# Signatures of the self-affinity of fracture and faulting in pre-seismic electromagnetic emissions.

S. M. Potirakis[a], K. Eftaxias[b], J. Kopanas[b], and G. Antonopoulos[c]

a. Department of Electronics, Technological Education Institute (TEI) of Piraeus, 250 Thivon & P. Ralli, GR-12244, Aigaleo, Athens, Greece, spoti@teipir.gr .

b. Department of Physics, Section of Solid State Physics, University of Athens, Panepistimiopolis, GR-15784, Zografos, Athens, Greece, ceftax@phys.uoa.gr .

c. Department of Environmental Technology and Ecology, Technological Education Institute (TEI) of the Ionian Islands, Panagoulas road, GR-29100, Zante, Greece, sv8rx@teiion.gr.

**Abstract**

Of particular interest is the detection of precursors of an impending rupture. Theoretical, numerical studies along with laboratory experiments indicate that precursory signs of an impending failure are the sudden drop of fractal dimension and entropy, along with the anticorrelated, for large system sizes, rising of Hurst exponent and drop of a frequency-size power-law scaling exponent. Based on the widely accepted concept of the self-affine nature of faulting and fracture, we examine whether these precursory signs exist in the fracto-electromagnetic emissions resulting from the activation of a single fault.

## I. INTRODUCTION

Understanding how materials break is a fundamental problem that has both theoretical and practical importance [1]. During the past two decades, considerable effort has been devoted by scientists to the study of damage and fracture in heterogeneous media (e.g., rocks) [2,3]. Of particular interest is the detection of precursors of an impending rupture. Laboratory studies have detected various precursory signatures of an impending failure [3-8]. Theoretical and numerical studies have devoted efforts for the explanation of the experimental precursory signs and suggested new ones [e.g., 9-13].

Earthquakes (EQs) are large-scale fracture phenomena in the Earth's heterogeneous crust. Since the early work of Mandelbrot [14], the self-affine nature of faulting and fracture is widely documented from the analysis of data both field observations and experiments [3 and references therein]. The question naturally arising is whether the precursory signs reported by laboratory, theoretical and numerical studies are also extended to the activation of a single fault. Herein we focus on this point, checking for compatibility with:

1. Theoretical studies performed by Lu et al. [15] found that the Fractal Dimension (FD) and entropy decreases as the damage in a disordered media evolves. A sudden drop of FD might be viewed as a likely precursor prior to a final catastrophic failure.

2. Long-range connective sandpile (LRCS) models [16-18] predict a negative correlation between Hurst exponent $H$ and a frequency-size power-law scaling exponent $B$ (or the FD) for large system sizes, which seem to be consistent with studies of earthquake fault systems and real seismicity data [17-20]. The $B-$ values (and FD) typically reduce prior to large avalanches, which mimics the observed precursory phenomena of the Gutenberg - Richter $b-$ values in real seismicity, while the $H-$ values increase.

3. A self-affine model (SAM) for the seismicity that mimics the fault friction by means of two fractional Brownian profiles that slide one over the other has been introduced by Hallgass et al. [20]. An earthquake occurs when there is an overlap of the two faces and its energy is assumed proportional to the overlap surface. The SAM exhibits the dependence of the Gutenberg-Richter law exponent to the roughness, $H$, of the fault surface profiles. More precisely, in their numerical simulations they observed that the probability of an earthquake releasing an energy $E$, $P(E)$, is following the power law $P(E) \propto E^{-1-\gamma}$, where $\gamma = 1 - H/(d-1)$ in the general $d-$ dimension case.

Finally, we check whether laboratory results are also compatible with the corresponding ones rooted in the activation of a single fault.

Crack propagation is the basic mechanism of material's failure. The motion of a crack has been shown to be governed by a dynamical instability causing oscillations in its velocity and structure on the fracture surface. Experimental evidence indicates that the instability mechanism is that of local branching: a multicrack state is formed by repetitive, frustrated microfracturing events [21].

Electromagnetic (EM) emissions in a wide frequency spectrum ranging from kHz to MHz are produced by cracks' opening, which can be considered as the so-called precursors of general fracture [22 and references therein]. The radiated EM precursors are detectable both at laboratory [5-8,23-25] and at geophysical scale [22,26-33]. An important feature at laboratory scale is that the MHz radiation precedes the kHz one: the kHz EM emission is launched in the tail of pre-fracture EM emission from 97% up to 100% of the corresponding failure strength [34 and references therein]. Clear fracture-induced MHz-kHz EM precursors have been detected over periods ranging from approximately a week to a few hours prior to significant EQs [22,28-34]. Importantly, the MHz radiation precedes the kHz one as it happens at the laboratory scale [22,28-34]. The remarkable asynchronous appearance of these precursors indicates that they refer to different stages of the EQ preparation process. The following two stage model of EQ generation by means of pre-fracture EM activities has been proposed [29,30]: (i) The pre-seismic MHz EM emission is thought to be due to the fracture of the highly heterogeneous system that surrounds the family of large high-strength entities distributed along the fault sustaining the system. It can be described as analogous to a thermal continuous phase transition while a Levy-walk-type mechanism can drive the heterogeneous system to criticality. (ii) The final kHz EM radiation is due to the fracture of the aforementioned large high-strength entities themselves. A sequence of kHz EM pulses is emerged where there is an intersection between the two rough profiles of the fault.

Since, according to this model, the kHz EM emissions are considered to stem from the last stage of the EQ preparation process, we are seeking for the above mentioned precursory signs in the kHz emissions. Our analysis is performed by means of: (i) FD evolution, estimated through the Hurst exponent resulting from rescaled-range ($R/S$) analysis, detrended fluctuation analysis (DFA), and spectral power law analysis; (ii) universal roughness of fracture surfaces; (iii) Gutenberg-Richter frequency-magnitude exponent $b$ evolution, calculated both directly and through the analysis in terms of a non-extensive model for earthquake dynamics and (iv) in terms of the recently proposed fuzzy entropy (FuzzyEn).

The results obtained after the analysis reveal good agreement to the corresponding theoretical, numerical and laboratory ones. The results indicate that the fracto-electromagnetic emissions associated with the activation of a single fault are compatible to the self-affine nature of fracture and faulting and provide clear indications that critical fracture is approaching.

The well documented fracture-induced kHz EM signal associated with the Athens EQ, with magnitude 5.9, occurred on 7 September 1999, e.g., [22,29-33], is employed in this contribution as a test case. Part of the recorded time series covering 11 days period from 28 August 1999, 00:00:00 (UT), to 7 September 1999, 23:59:59 (UT), and containing the candidate precursor signal is shown in Fig. 1.

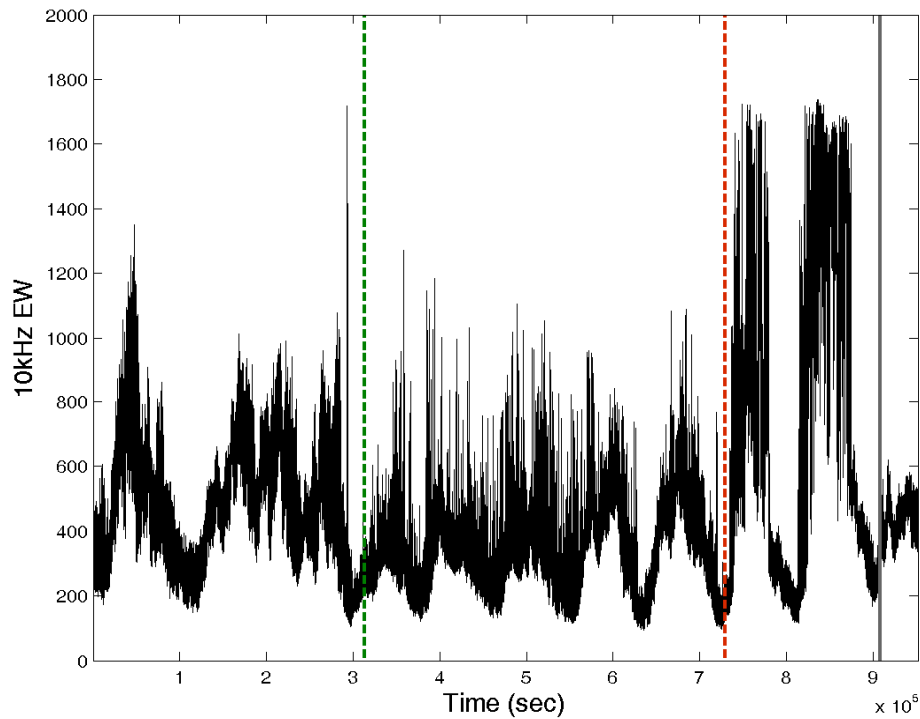

**FIG. 1** Part of the recorded time series of the 10 kHz (East–West) magnetic field strength (in arbitrary units) covering 11 days period from 28 August 1999, 00:00:00 (UT), to 7 September 1999, 23:59:59 (UT), associated with the Athens EQ. The vertical solid grey line indicates the time of the Athens EQ occurrence. The (left) vertical broken green line roughly indicates the start of the candidate precursor. The (right) vertical broken red line indicates where the damage evolution of the fault approaches the critical point.

The (left) vertical broken green line in Fig. 1 roughly indicates the start of the candidate precursor. Before that the recordings correspond to the background electromagnetic noise at the position of the station. The (right) vertical broken red line in Fig. 1 indicates where the damage evolution of the fault approaches the critical point.

## II EXPERIMENTAL DATA ANALYSIS

In the following we examine whether precursory characteristics predicted by theoretical, numerical studies and laboratory experiments are also included in fracture-induced EM emissions. First, we analyze the experimental data in terms of the Hurst exponent and the corresponding FD temporal evolution, by means of the rescaled-range ($R/S$) analysis, detrended fluctuation analysis (DFA), and spectral power law analysis. Then, we study the evolution of Gutenberg-Richter frequency-magnitude exponent, calculated both directly and through the analysis in terms of a non-extensive model for earthquake dynamics. Finally, the entropy evolution is examined by means of the FuzzyEn.

### A. Hurst exponent

The rescaled range $R/S$ analysis was chosen for the direct calculation of the Hurst exponent, while the DFA method, as well as the spectral power law method, were chosen for the indirect estimation of Hurst exponent under the fractional Brownian motion (fBm) model hypothesis, which is respectively checked for its validity.

The exponent $H$ characterizes the persistent / anti-persistent properties. The range $0 < H < 0.5$ indicates anti-persistency, which means that if the fluctuations presently increase, it is expected to change tendency in near future (negative feedback mechanism). On the contrary, persistent behavior is characterized by $0.5 < H < 1$ and then the underlying dynamics is governed by a positive feedback mechanism. The employed methods for the calculation / estimation of $H$ are briefly reviewed in the following.

The $R/S$ analysis [35,36] is based on two quantities: first, the range $R_n$, which is the difference between the maximum and minimum values of the accumulated departure of the time series from the mean, calculated over each one ($n = 1, 2, ..., d$) of the $m$−samples long sub-series in which the time-series can be divided, and second, the standard deviation of the corresponding sub-series $S_n$. The so-called *rescaled range* is exactly the ratio of $R$ by $S$. Hurst found that ($R/S$) scales by power - law as time (i.e., the sample length $m$ of the sub-series) increases,

$$(R/S)_m \propto m^H, \qquad (1)$$

where $H$ is the Hurst exponent. The exponent $H$ is estimated as the linear slope of a $\log(R/S)_m - \log m$ representation.

Detrended Fluctuation Analysis (DFA) is a straightforward technique for identifying the extent of fractal self-similarity in a seemingly non-stationary time-series [37,38]. After dividing a time-series to sub-series of $m-$samples length, the root mean-square fluctuation for the integrated and detrended series, $F(m)$ is calculated. Repeating this calculation for different $m$, a power-law relation between $F(m)$ and time (expressed by sub-series length $m$)

$$F(m) \propto m^a \qquad (2)$$

indicates the presence of scaling. The DFA exponent $a$ is estimated as the linear slope of a $\log F(m) - \log m$ representation.

Moreover, if an observed time-series is a temporal fractal, it should follow a spectral power law

$$S(f) \propto f^{-\beta}, \qquad (3)$$

where $S(f)$ is the power spectral density, and $f$ the frequency. The spectral power law exponent is estimated as the linear spectral slope $-\beta$ of a $\log S(f) - \log f$ representation of the power spectrum. The quality of fit to spectral power-law (as well as for the power laws of the other two methods) is usually measured in terms of the linear correlation coefficient, $r^2$.

The spectral scaling exponent $\beta$ is related to the Hurst exponent, $H$:

$$\beta = 2H + 1, \qquad (4)$$

with $0 < H < 1$ ($1 < \beta < 3$) for the fBm model [39].

Moreover, the relations between the DFA exponent $a$, the Hurst exponent $H$, and the spectral power law exponent $\beta$ in the case of an fBm time-series are [40,41]

$$H = a - 1 \qquad (5)$$

and $$\beta = 2a - 1. \qquad (6)$$

The Hurst exponent is first directly calculated using the $R/S$ method, and the result is simply denoted by $H$ in the following. Then the Hurst exponent is also estimated, under the fBm hypothesis, from the calculated DFA exponent $a$ (using Eq. 5) and the spectral scaling exponent $\beta$ (using Eq. 4), while the estimated Hurst exponents are denoted in the following as $H_a$ and $H_\beta$, respectively, in order to be easily discriminated from the directly calculated, by the $R/S$ method, $H$.

The $R/S$ method Hurst exponent, $H$, and the DFA exponent, $a$, were calculated using on successive non-overlapping 1024 samples long windows, and running time average of four windows with 25% overlapping. Only the exponent values which arose for fitting of correlation coefficient $r^2 > 0.85$ to the corresponding power laws were considered here. The resulting, $H$ and $H_a$ are depicted in Figs 2b and 2c, respectively.

The spectral scaling exponent $\beta$ was estimated by calculating the morlet wavelet spectrum on successive, overlapping, time-windows of 1024 samples width each, an overlap of 75%, i.e., sliding with a step of 256 samples, and running time average of four windows with 25% overlapping. Only the $\beta$ exponent values which presented correlation coefficient $r^2 > 0.85$ were considered here. The results for the corresponding estimated $H_\beta$, resulting from $\beta$ supposing an fBm model and therefore employing Eq. (4), are presented in Fig. 2d.

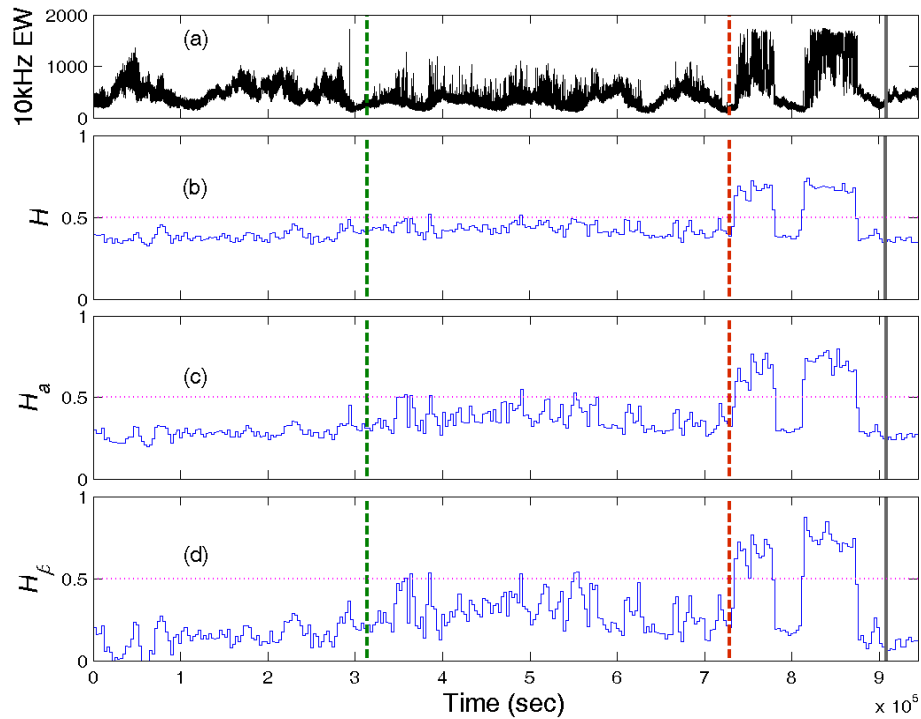

**FIG. 2** (a) Part of the recorded time series of the 10 kHz (East–West) magnetic field strength (in arbitrary units) covering 11 days period from 28 August 1999, 00:00:00 (UT), to 7 September 1999, 23:59:59 (UT), associated with the Athens EQ. The corresponding variation vs. time of Hurst exponent, (b) $H$, resulting from $R/S$ method, (c) $H_a$, estimated via DFA, and (d) $H_\beta$ calculated from spectral power law. The common horizontal axis is the time (in s), denoting the relative time position from the beginning of the analyzed part of the EM recording. The vertical lines have the same position and meaning as in Fig. 1. (For interpretation of the references to color in this figure, the reader is referred to the online version of this article.)

First of all we focus on the validity of the fBm model hypothesis. We claim that the above results verify the validity of the fBm model hypothesis for the signal under analysis for the following two reasons: (i) the comparison among the differently calculated/ estimated Hurst exponents reveal that all of them are consistent to each other. Note that, the estimated values are very much the same throughout the signal duration, i.e., to the right of the first (green) broken line; (ii) the calculated spectral scaling exponent values are within the frame of the expected, for fBm time-series, $1 < \beta < 3$ [39]. If the fBm hypothesis wasn't valid, then at least one of the two independent indirect methods for the estimation of the Hurst exponent, under the fBm hypothesis, should lead to different results from those obtained through its direct calculation by the $R/S$ method. We note that during the last part of the analyzed time-series, following the right (red) vertical broken line, the two strong EM bursts follow the persistent ($H > 0.5$) fBm model.

Fracture surfaces have been found to be self-affine following the fractional Brownian motion (fBm) model over a wide range of length scales. Specifically, they are characterized by $H \sim 0.7 - 0.8$, weekly dependent on the failure mode and the nature of the material, leading to the interpretation that this range of Hurst exponent values constitute a universal indicator of surface fracture [42-47].

According to our two-stage model (see Sec. I), the kHz EM radiation is due to the interaction of the two rough profiles of the fault. Therefore, it is expected that the roughness of the analyzed kHz time-series, should be consistent to the global values of fault roughness. Indeed, all the estimated Hurst exponent values during the two strong EM bursts converge to $H \sim 0.7$.

As recently pointed out in Chen et al. [17] and Lee et al. [18], Hallgass et al. [20] have introduced a self-affine model (SAM) for the seismicity that mimics the fault friction by means of two fractional Brownian profiles that slide one over the other. Since the roughness index, $H$, of the analyzed EM time series is $H \sim 0.7$, the SAM predicts that the probability an EM pulse having an energy $E$ should be denoted by $P(E) \propto E^{-1-\gamma}$, where $\gamma = 1 - H = 0.3$ and thus $P(E) \propto E^{-1.3}$, given that the EM time-series is a two-dimensional variation, i.e. $d = 2$. The question arises whether the energy of EM pulses follow the power law $P(E) \propto E^{-1.3}$. Indeed, the cumulative distribution function of the specific EM time-series amplitudes has been proved to follow the power law $N(>A) \sim A^{-0.62}$ [48], and, consequently, the distribution function of the energies follows the power-law $P(E) \sim E^{-1.31}$ [49]. It is noted that Petri et al. [50] found a power-law scaling behavior in the acoustic emission energy distribution with $-1-\gamma = -1.3 \pm 0.1$. Houle and Sethna [51] found that the crumpling of paper generates acoustic pulses with a power-law distribution in energy with $-1-\gamma \in (-1.6, -1.3)$. On the other hand, Cowie et al. [52], Sornette et al. [53], and Cowie et al. [54], have developed a model of self-organized EQs occurring on self-organized faults. Their theoretical study suggests that the corresponding exponent value should be $-1-\gamma = -1.3$.

Finally, a physical modeling of the formation and evolution of seismically active fault zones has been studied in the frame of laboratory experiments which also ended-up to compatible values of Hurst exponents ($H \sim 0.7$) for both space and time analysis [4], which is also in agreement with the obtained results for the kHz EM time-series.

The above results on Hurst exponent are consistent both to the numerical results for the LRCS model [16-18], predicting increase of Hurst exponent prior to large events, and the SAM model [20], yielding an energy distribution exponent very close to the predicted by the model and past laboratory experiments.

## B. Fractal dimension

Given the validity of the fBm model hypothesis, the Hausdorff-Besicovitch FD $D_h$ can be estimated from the relation [39,55]

$$D_h = 2 - H = (5-\beta)/2. \qquad (7)$$

The corresponding FD values resulting from the directly calculated ($R/S$) Hurst exponent, $H$, and the estimated from the calculated DFA exponent $a$, $H_a$, and the spectral scaling exponent $\beta$, $H_\beta$, are in that order denoted by $D_h$, $D_{ha}$ and $D_{h\beta}$. The analysis results are depicted in Figs 3b, 3c and 3d, respectively.

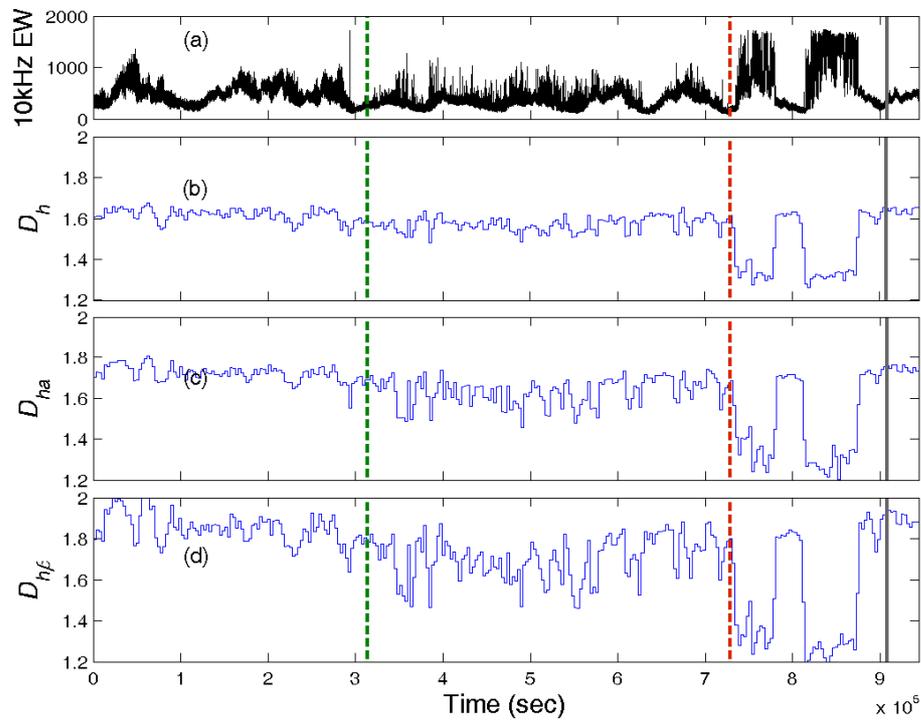

**FIG. 3** (a) Part of the recorded time series of the 10 kHz (East–West) magnetic field strength (in arbitrary units) covering 11 days period from 28 August 1999, 00:00:00 (UT), to 7 September 1999, 23:59:59 (UT), associated with the Athens EQ. The corresponding temporal variation of the Hausdorff-Besicovitch FD (b) $D_h$, as resulting from $H$ (c) $D_{ha}$ as resulting from $H_a$, and (d) $D_{h\beta}$ calculated from $H_\beta$, all calculated using Eq. (7). The common horizontal axis is the time (in s), denoting the relative time position from the beginning of the analyzed part of the EM recording. The vertical lines have the same position and meaning as in Fig. 1. (For interpretation of the references to color in this figure, the reader is referred to the online version of this article.)

From Fig 3 it is apparent that all the estimated FDs suddenly drop during the two strong EM pulses, compared to the background noise as well as to the first part of the signal (the part between the green –left– and red –right– broken vertical lines). Note that all of them reach values of $D_h \sim 1.3$ during these two strong pre-EQ EM emissions. These results are consistent to theoretical findings of Lu et al. [15], predicting sudden drop of FD prior to the final catastrophic failure. Moreover, the sudden drop of FD has also been observed in laboratory experiments [4], while the FD value yielded for the EM signal under analysis is also consistent to the corresponding of the geophysical scale, obtained for distribution of rupture fault lengths irrespective of their positions [48,56]. This, for a single major fault, has been estimated to $D \sim 1.2$ by seismological measurements as well as theoretical studies [e.g. 57,58, and references therein].

The direct calculation of the FD for the signal of Fig. 1 has already been presented in [59], in terms of the box-counting and the Higuchi's algorithms, giving a FD $\sim 1.6$ during the two strong EM pulses. These two methods are also successfully highlighting the sudden drop of fractal dimension prior to the final catastrophic failure in agreement to the theoretical findings of Lu et al. [15].

Nevertheless, we observe a difference between the results of [59] and the indirectly estimated FD of Fig. 3 as a result of the sensitivity of the box-counting and the Higuchi's algorithms to noise. It is a common knowledge that all practical FD estimates are very sensitive to numerical or experimental noise, e.g., [60]. The presence of noise leads to FD estimates which are higher than the actual FD. Since the field-acquired EM time series under analysis is certainly contaminated by measurement noise, the calculated FD values were higher than the actual, while the different sensitivity to the measurement noise led to differences between the FD estimates through the two algorithms employed in [59]. Although various algorithms for calculating FD have been developed, a general solution is not available. It is often said, e.g., [61], that at least two different algorithms are needed for a faithful representation of the FD of a time series.

However, the indirect estimation presented here (Fig. 3) through three different methods led to similar results. Therefore, provided the proven validity of the considered fBm model, as well as the consistency of the resultant Hurst exponent values with the corresponding numerical model and laboratory facts, one could end-up to the conclusion that the Hurst exponent sourced FDs have to be considered more reliable than the ones calculated by the Higuchi or the box-counting methods, which are probably prone to higher FD values than the actual ones due to measurement noise.

### C. Frequency-size law

Earthquake dynamics have been found to follow the frequency-magnitude scaling relation, known as Gutenberg - Richter law [62]

$$\log N(>M) \sim -bM, \qquad (8)$$

where $N(>M)$ is the number of earthquakes with magnitude greater than $M$ occurring in a specified area and time and the coefficient $b$, called "the $b$-value", is the negative slope of $\log N(>M)$ vs. $M$ diagram.

A model for EQ dynamics based on a non-extensive Tsallis formalism, starting from fundamental principles, has been recently introduced by Sotolongo-Costa and Posadas [63] and revised by Silva et al. [64]. This approach leads to a non-extensive Gutenberg–Richter type law for the magnitude distribution of EQs:

$$\log\left[N(>M)\right] = \log N + \left(\frac{2-q}{1-q}\right)\log\left[1 - \left(\frac{1-q}{2-q}\right)\left(\frac{10^{2M}}{a^{2/3}}\right)\right], \quad (9)$$

where $N$ is the total number of EQs, $N(>M)$ the number of EQs with magnitude larger than $M$, $M \sim \log\varepsilon$. $a$ is the constant of proportionality between the EQ energy, $\varepsilon$, and the size of fragment, $r$, $(\varepsilon \sim r^3)$. It is reminded that the entropic index $q$ characterizes the degree of non-extensivity. Importantly, the associated with Eq. (9) $q$-values for different regions (faults) in the world are restricted in the region 1.6 – 1.7 [64].

The $q$-parameter included in the non-extensive formula of Eq. (9) is associated with the $b$-value by the relation [65]:

$$b_{est} = 2 \cdot \frac{2-q}{q-1} \quad (10)$$

In order to further verify the compliance of the analyzed EM recordings to the LRCS model, we check whether the $B$-values typically reduce prior to large avalanches while the $H$-values increase. Towards this direction the Gutenberg - Richter law and its non-extensive variant were employed. Both of them were applied using the notion of fracto-electromagnetic emission event, or "electromagnetic earthquake" (EM-EQ), within the frame of the self-affine nature of fracture and faulting. Within this frame, a fracto-electromagnetic emission event is considered to correspond to a fracture event which is regarded as analogous to an EQ at the geophysical scale. If $A(t_i)$ refers to the amplitude of the pre-EQ EM time-series, we regard as amplitude of a candidate "fracto-electromagnetic emission" the difference $A_{fem}(t_i) = A(t_i) - A_{noise}$, where $A_{noise}$ is the maximum value of the EM recording during a quiet period, namely far from the time of the EQ occurrence. We consider that a sequence of $k$ successively emerged "fracto-electromagnetic emissions" $A_{fem}(t_i)$, $i = 1,\ldots,k$ represents the EM energy released, $\varepsilon$, during the damage of a fragment. We shall refer to this as an "electromagnetic earthquake" (EM-EQ). Since the sum of the squared amplitude of the fracto-electromagnetic emissions is proportional to their energy, the magnitude $M$ of the candidate EM-EQ is given by the relation $M \sim \log\varepsilon = \log\left(\sum\left[A_{fem}(t_i)\right]^2\right)$.

Both frequency-size laws were fitted in the time domain, on three large parts of the signal in order to ensure adequate statistics for the analysis. For the Gutenberg - Richter law, the fitting had a correlation coefficient $r^2 > 0.99$ for all three cases, while for its non-extensive variant a fitting error <1% was achieved.

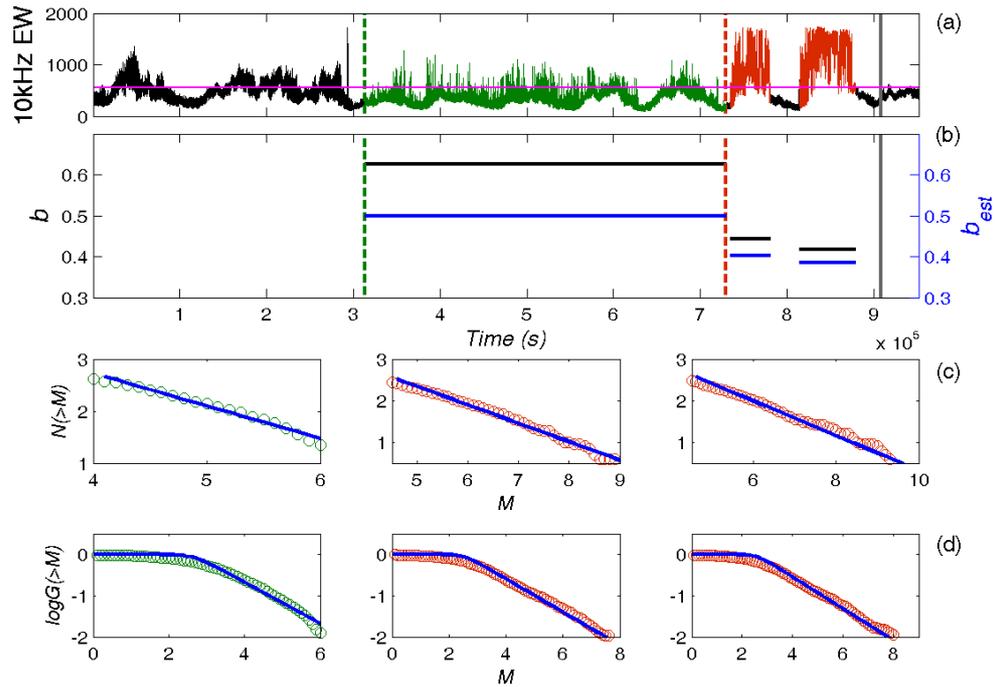

**FIG. 4** (a) Part of the recorded time series of the 10 kHz (East–West) magnetic field strength (in arbitrary units) covering 11 days period from 28 August 1999, 00:00:00 (UT), to 7 September 1999, 23:59:59 (UT), associated with the Athens EQ. (b) Temporal variation of the $b-$value, and the $b_{est}$, estimated from the non-extensive $q$ parameter (Eq. (10)). (c) Fitting of the Gutenberg-Richter law and (d) the non-extensive Gutenberg-Richter law, on the three parts of the analyzed signal (color and position correspondence from left to right) The common horizontal axis is the time (in s), denoting the relative time position from the beginning of the analyzed part of the EM recording. The magenta horizontal line on Fig. 4(a) indicates the noise level threshold $A_{noise} = 620 a.u.$. The vertical lines have the same position and meaning as in Fig. 1. (For interpretation of the references to color in this figure, the reader is referred to the online version of this article.)

From Fig. 4 it is apparent that both methods reveal a sudden drop of $b-$value during the two strong EM emissions, during which Hurst exponent was suddenly raised. Therefore, the analyzed EM recordings could be said to be compliant to the LRCS model. The lowering of the corresponding $b-$values indicates the increase of the number of large events against the number of small ones.

The sudden reduction of $b-$value is a scale-invariant precursor of an impending rupture. Indeed, during the deformation of rock in laboratory experiments, small cracking events emerge which radiate elastic waves in a manner similar to EQ [4,66]. These emissions were found to obey the Gutenberg-Richter type relation. Acoustic Emissions (AE) from rock fracturing present a significant fall of the observed $b-$values as the main event approaches, i.e., indicate a significant decrease in the level of the observed $b-$values immediately before the critical point, e.g., [4,66-70]. The sudden reduction of the $b-$value before the EQ occurrence is also reported at seismicity scale by several researchers, e.g., [15,71-73]. Moreover, it is widely known that FD is directly proportional to the $b-$value [15]. Therefore, a sudden reduction of FD and $b-$value is observed at all three scales (laboratory, fault, seismicity).

### D. Revealed physical pictures in view of the scalogram

The physical pictures outlined by the presented results are further enhanced by the time-scale analysis of the pre-EQ signal. The morlet scalogram of the signal under analysis is depicted in Fig. 5. It has to be mentioned that although a morlet wavelet was used, on the basis of its popularity, for the presented scalogram, nine mother wavelet cases were investigated, namely: morlet, meyer, mexican hat, haar, as well as different orders of coiflets, daubechies and symlets wavelets, all of them resulting to very much the same time-scale representation.

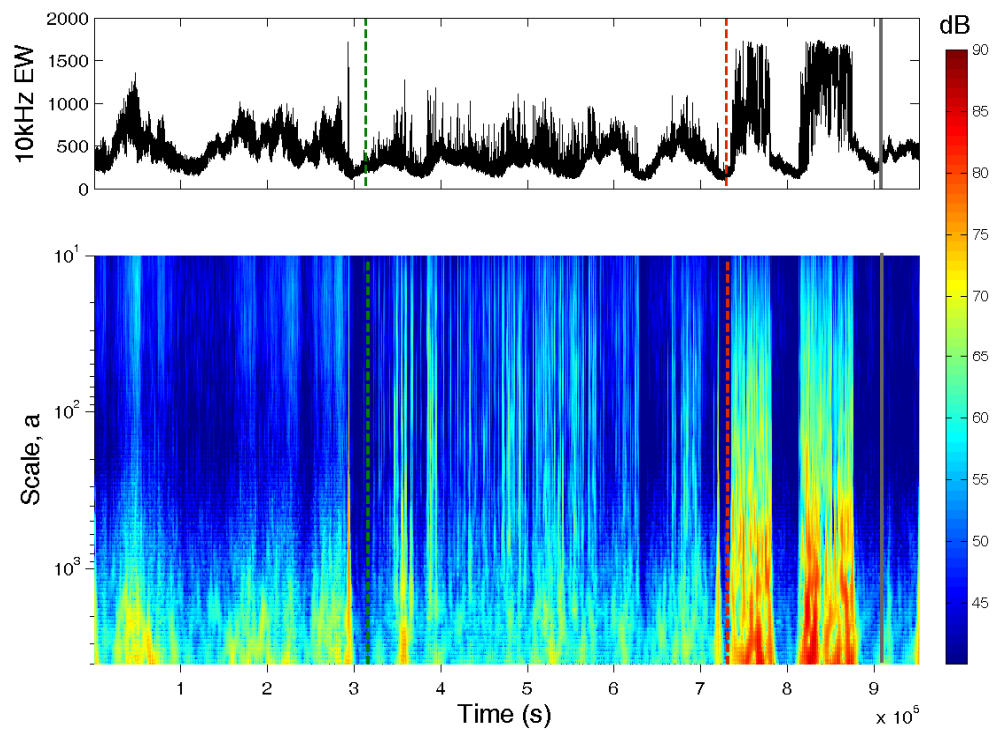

**FIG. 5** (a) Part of the recorded time series of the 10 kHz (East–West) magnetic field strength (in arbitrary units) covering 11 days period from 28 August 1999, 00:00:00 (UT), to 7 September 1999, 23:59:59 (UT), associated with the Athens EQ. (b) The corresponding morlet wavelet scalogram, with the vertical axis corresponding to the scale, $a$, of the wavelet (time scale, reciprocal to the wavelet "frequency") and the color representing the power spectral level in dB (side colorbar). The common horizontal axis is the time (in s), denoting the relative time position from the beginning of the analyzed part of the EM recording. The vertical lines have the same position and meaning as in Fig. 1. (For interpretation of the references to color in this figure, the reader is referred to the online version of this article.)

We observe on Fig. 5 that during the two strong EM bursts, the EM emission extends to all scales. However, the higher part of the emitted energy is localized to the lower frequencies. Therefore, one could conclude that the underlying fracture phenomenon extends to all fracture scales in a coherent way but with a preference to large scale fractures. This is in agreement to the physical picture resulting from the temporal evolution of the $\beta$ exponent values, as indicated by the temporal evolution of Hurst exponent Fig 2d, taking into account Eq. (4). The $\beta$ exponent is shifted to higher values in the tail of the EM emission indicating the formation of a long-term memory in the system. Moreover, the FD values resulted from the Hurst exponent estimations are not only compatible with the fBm model of fracture but also with the increase of $\beta$, the reduction of $b$-value, and all the above physical pictures. Indeed, these FD values are interrelated to Hurst exponent and $\beta$ within the frame of fBm model (Eqs 4-7), while the verified anticorrelated relation between $H$-exponent and $b$-value according to the LRCS model provides a linkage between FD and $b$-value behavior. The physical picture behind this linkage is the following: the sudden reduction of the FD values observed during the two strong EM bursts implies the sudden domination of large events over the small ones since this is expected to match up to a more incomplete "fill" of space (larger entities leave more space between them than smaller ones) and it is reflected to the lower filling capacity (lower FD) of the corresponding time-series on the amplitude-time plane.

All the above physical pictures are compatible to the final stage of the activation of a fault.

### E. Fuzzy entropy

Fuzzy entropy (FuzzyEn) [74,75], like its ancestors ApEn and SampleEn [75], is a "regularity statistics" that quantifies the unpredictability of fluctuations in a time series. For the calculation of FuzzyEn, vectors' similarity is defined by fuzzy similarity degree based on fuzzy membership functions and vectors' shapes. The gradual and continuous boundaries of fuzzy membership functions lead to a series of advantages like the continuity as well as the validity of FuzzyEn at small parameters, higher accuracy, stronger relative consistency and less dependence on data length. FuzzyEn can be considered as an upgraded alternative of SampEn (and ApEn) for the evaluation of complexity, especially for short time series contaminated by noise.

FuzzyEn calculations were performed according to the algorithm provided in [75], on successive non-overlapping 1024 samples long windows, and running time average of four windows with 25% overlapping. It is noted that for the calculation of FuzzyEn, the exponential function has been used as the fuzzy membership function, $\mu\left(d_{ij}^{m}, r\right) = \exp\left(-\left(d_{ij}^{m}/r\right)^{n}\right)$, with $n=2$, for $m=2$ and $r=0.65 \cdot STD$, where $STD$ is the standard deviation of the analyzed time-series fragment, allowing fragments with different amplitudes to be compared.

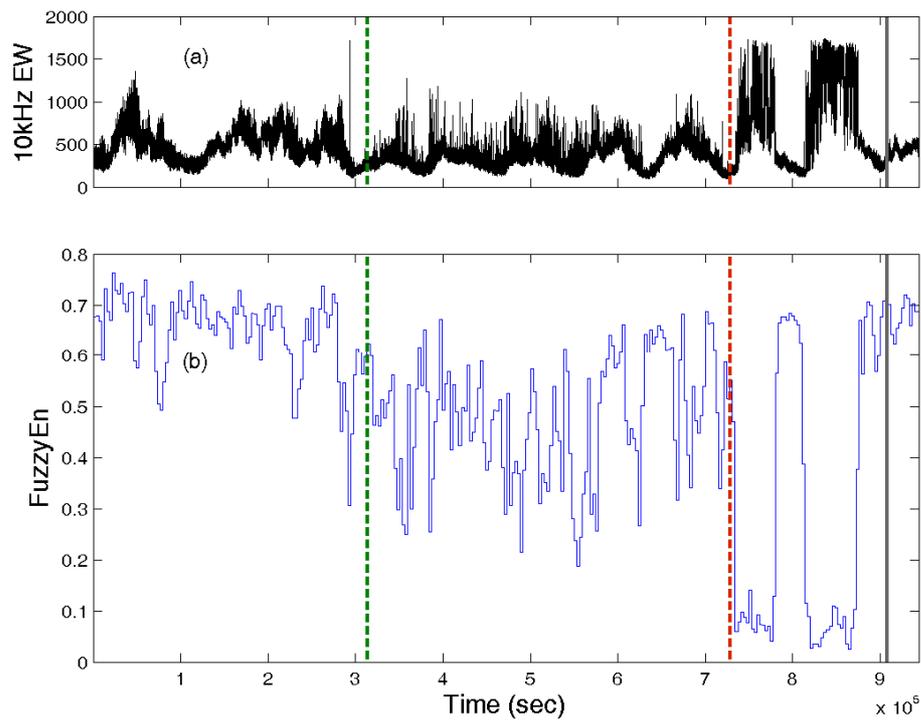

**FIG. 6** (a) Part of the recorded time series of the 10 kHz (East–West) magnetic field strength (in arbitrary units) covering 11 days period from 28 August 1999, 00:00:00 (UT), to 7 September 1999, 23:59:59 (UT), associated with the Athens EQ. (b) The corresponding FuzzyEn. The common horizontal axis is the time (in s), denoting the relative time position from the beginning of the analyzed part of the EM recording. The vertical lines have the same position and meaning as in Fig. 1. (For interpretation of the references to color in this figure, the reader is referred to the online version of this article.)

From Fig. 6 it can be observed that lower entropy values, compared to that of the background noise, can be observed between the green (left) and the red (right) broken vertical lines, although sparsely distributed in time. On the other hand, the entropy values suddenly drop during the two strong EM pulses signifying a different behavior, a new distinct phase in the tail of the EQ preparation process which is characterized by a significantly higher degree of organization and lower complexity in comparison to that of the preceding phase.

This final phase of precursory EM phenomenon combines a sudden drop of entropy and a sudden drop of FD (see Fig. 3), a combination of precursory signs which have been reported by Lu et al. [15] as a quantitative measure of the damage localization (or the clustering degree of microcracks/voids), and a likely precursor prior to a final catastrophic failure.

## III CONCLUSIONS

In this contribution, we focused on the sudden drop of fractal dimension and entropy, along with the anticorrelated, for large system sizes, rising of Hurst exponent and drop of a frequency-size power-law scaling exponent. These have been indicated as precursory signs of an impending failure by theoretical, numerical studies along with laboratory experiments. We analyzed fracto-electromagnetic emissions resulting from the activation of a single fault proving that all these signs are included in these emissions, further supporting the concept of the self-affine nature of faulting and fracture.